\documentclass[aps,prl,twocolumn,groupedaddress]{revtex4}

\usepackage{graphicx}

\begin{document}
\title{Simplicity and scaling - size of a real polymer in three (or any) dimensions}
\author{C.P. Lowe}
\affiliation{van 't Hoff Institute for Molecular Science, University of Amsterdam, Nieuwe Achtergracht 166, 1018 WV Amsterdam,
The Netherlands}
\author{M.W. Dreischor}
\affiliation{Toegepaste Industri\"{e}le Procesbeheersing, Kruislaan 419, 1098 VA Amsterdam,
The Netherlands}
\begin{abstract}
We examine the scaling of the linear dimension of the system size of a real polymer
solution at constant excess free energy and in two different spacial dimensionalities, $d=d_0$ and $d=d_1$. 
Standard results for the functional form of the excess free energy
lead to the conclusion that the scaling exponent $\nu(d)$ satisfies 
\mbox{$\nu(d_0) - \nu(d_1) = d_0^{-1} - d_1^{-1}$}. Taking the
critical dimensionality as a point of reference ($\nu(4)=1/2$)
gives a scaling exponent $\nu(d) = 1/4 +1/d$, in agreement with the accepted result
for two-dimensions ($\nu(2) = 3/4)$ and the first term in the epsilon ($d-4$) expansion.
For the unsolved case of three dimensions it predicts $\nu(3)=7/12$. Several
simplifying features of this result are pointed out.
\end{abstract}

\maketitle

An amazing thing about polymers is that, although they are
complex molecules, just a few parameters usually 
describe their macroscopic properties~\cite{deGennes1979}. For example, the polymer size, $L_p$, 
is given by the scaling relation  $L_p \sim l_0 n^{\nu(d)}$.
Here, the microscopic quantities are the number of monomers $n$, 
and a length $l_0$. The precise definition of the latter we discuss later.
The scaling exponent $\nu(d)$ generally depends on the
spacial dimensionality $d$. If the monomers do not
interact the chain is ``ideal'' and $L \sim b n^{\frac{1}{2}}$, independent
of the dimensionality.
Here $b$ is the Kuhn length (the root mean square separation between
adjacent monomers). More generally a polymer is ``real'', meaning that
the monomers do interact. Flory pointed out that this changes the
scaling~\cite{Flory1949}. Using a very simple model, he postulated that if the
interaction between monomers is repulsive then, $L_p \sim l_0 n^{\frac{3}{2+d}} (1 \leq d < 4)$
and $L_p \sim l_0 n^{\frac{1}{2}} (d \geq 4)$. So, for $d < 4$ the 
real chain is expanded relative to the ideal chain. For $d \geq 4$ 
the chain scales in the same way as an ideal chain, so the interactions
have no effect. In this respect Flory's model is correct.
For two dimensions it predicts $\nu(2) = 3/4$. Again, this is accepted as
correct~\cite{Nienhuis1982}. Nonetheless, it is regarded
as inexact for the most important case~-~namely three dimensions.

The reason for this is as follows.
De Gennes famously showed the equivalence of the real polymer model to the
$n=0$ limit of the n-vector model~\cite{deGennes1972}.
First, this shows that the real polymer problem has significance
beyond the realms of polymer physics. 
A nice example is that modelling the universe as a real polymer resolves
Olbers' paradox without assuming an expanding universe~\cite{Diao2007}.
Second, having identified this equivalence de Gennes could use
Wilson's method~\cite{Wilson1972} to show that
near to the critical dimensionality ($d=4$),
$L_p(\epsilon \rightarrow 0) \sim  n^{\frac{1}{2}+\frac{\epsilon}{16}}$,
where $\epsilon = 4 - d$.
In this limit (which we refer to as $d=3.99$), the Flory 
result gives $L_p( \epsilon \rightarrow 0 ) \sim l_0 n^{\frac{1}{2}+\frac{\epsilon}{12}}$.
Consequently, the prediction that for $d=3$, $L_p \sim l_0 n^{\frac{3}{5}}$
is approximate. Current numerical estimates are that actually $\nu(3)=0.5877 (\pm 0.0007)$~\cite{Sokal1995}.
So the exact value of this fundamental exponent,
almost the polymer equivalent of $\pi$, remains unknown.
Here we use the functional form of the excess free energy of a polymer solution in
different spacial dimensions to argue that in three dimensions $\nu(3)=7/12(=0.5833)$. We also point out
a number of simplifications
regarding the physics of polymers and polymer solutions that would then follow.

There are two microscopic lengths
involved in the problem, $b$ and $l$, where $l$ is the range of the monomer-monomer potential.
For a given value of the ratio $l/b$, the size of the chain can be written in terms
of either $l$ or $b$. Here we take the former. Where
proportionalities and similarities are used, from now on it is because we neglect 
constants dependent only on the dimensionality.
Consider the dimensionless excess free energy of a polymer solution, $\tilde{G}^*$,
where $\tilde{G}^* = \tilde{G}/kT$ with $\tilde{G}$ the excess free energy, $T$ the (constant) temperature,
and $k$ Boltzmann's constant. According to scaling theory~\cite{DoiEdwards1986} this takes the form
$\tilde{G}^* = N_p f_d(\phi_p)$,
where $f_d$ is some dimensionally dependent function, $N_p$ is the number of polymers in the system and $\phi_p$ is the ``polymer space
fraction''
\begin{equation}
\phi_p \sim \frac{N_p L_p^d}{L^d}
\end{equation}
Here $L_p$ is the linear dimension of the polymer in the dilute limit. In
the scaling ($n \rightarrow \infty$) limit this has the form
\begin{equation}
L_p(d) \sim b n^{\nu(d)}
\end{equation}
Equally well, we can write the dimensionless excess free energy in terms of
a dimensionally dependent function $g_d$ and the ``polymer length fraction'', as $\tilde{G}^* = N_p g_d(\Phi_p)$ where
\begin{equation}
\Phi_p \sim \frac{N_p^{1/d} L_p}{L}
\end{equation}
Because this parameter involves lengths, which can be compared in different dimensionalities,
it is more convenient for our purposes here. We now consider two systems in two different dimensionalities $d_0$ and $d_1$.
The Kuhn length $b$ in dimensionality $d_0$ is proportional to the Kuhn length in dimensionality $d_1$,
so there is no change in the relative size of the polymers in the two dimensionalities
because of a varying ratio of Kuhn lengths.
We now consider how the linear dimension of the system changes as we increase the number
of the polymers and/or monomers, such that the length fraction remains unchanged. That is,
such that $\tilde{G}^*(d_0) / \tilde{G}(d_1)$ depends on the dimensionality but
nothing else. To do this requires that
\begin{eqnarray}
L(d_0) &\propto& N_p^{1/d_0}n^{\nu(d_0)} b\\
L(d_1) &\propto& N_p^{1/d_1}n^{\nu(d_1)} b 
\end{eqnarray}
So the ratio of the system sizes in the two dimensionalities scales as 
\begin{equation}
\label{c_low}
\frac{L(d_0)}{L(d_1)} \propto N_p^{1/d_0 - 1/d_1} n^{\nu(d_0) - \nu(d_1)}
\end{equation}
The above form of the excess free energy also holds 
above the ``overlap concentration'', $\Phi_p >> 1$, where the polymers
strongly overlap and the solution looks like a monomer soup. In this limit
one expects that $n$ is irrelevant and only the monomer length fraction, $\Phi_m$, 
\begin{equation}
\Phi_m \sim \frac{ N_m^{1/d} b }{L}
\end{equation}
is relevant. Here $N_m(=n N_p)$ is the total number of monomers in the system.
In this limit the dimensionless excess free energy, in terms of a dimensionally
dependent function $h_d$, then takes the form $\tilde{G}^* = N_m h_d( \Phi_m )$.
Far above the overlap concentration, but still in the semi-dilute
regime ($\Phi_p >> 1$, $\Phi_m << 1$) both functional forms of the
excess free energy are valid, that is $g_d(\Phi_p) \sim n h_d(\Phi_m)$.
This leads to generally accepted scaling of the free energy in this limit~\cite{deGennes1979}.
Specifically, in terms of $\Phi_m$, $\tilde{G}^*= N_p g_d ( \Phi_m n^{\nu(d)-1/d)})$, so for the
above to be true requires \mbox{$\tilde{G}^*/N_p \sim \Phi_p^{d/(d\nu(d)-1)}$}. 
We should note, however, that for one dimension, where $\nu(1) =1 $, the dimensionless free energy is a function
of only $\Phi_m$ so both functional forms 
for the excess free energy cannot be satisfied simultaneously. This is
reflected in the nonsensical prediction 
that in one dimension the free energy is infinite.

Now we can again consider the 
transformation in dimensionalities $d_0$ and $d_1$ (neither of which is now unity) described above.
In terms of the monomer length fraction we have
\begin{eqnarray}
L(d_0) & \propto & {\left( N_p n \right) }^{1/d_0} b\\
L(d_1) & \propto & {\left( N_p n \right) }^{1/d_1} b 
\end{eqnarray}
so the ratio of the system sizes scales as
\begin{equation}
\label{c_high}
\frac{L(d_0)}{L(d_1)} \propto {\left( N_p n \right)}^{1/d_0 - 1/d_1} 
\end{equation}
In the semi-dilute limit, both~\ref{c_low} and~\ref{c_high} are true,
implying that
\begin{equation}
\label{d_depend}
 \nu(d_0) - \nu(d_1) = 1/d_0 - 1/d_1
\end{equation}
If we take
four dimensions to define $d_0$ (that is $d_0 = 4, \nu(4) = 1/2$),
then, because $d_1$ can be any other dimensionality $d$, we have
\begin{equation}
\label{nu}
\nu(d) = \frac{1}{d} + \frac{1}{4}
\end{equation}
For two dimensions this yields $\nu(2) = 3/4$, the accepted result.
For three dimensions it yields $\nu(3) = 7/12$. Note that we
could equally well take the two dimensional result as a point of reference.
In this case we would recover the four dimensional
result. According to equation~\ref{nu} the excess free energy 
takes a particularly simple form above the overlap concentration, $\tilde{G}^*\sim N_p \Phi_p^{4}$,
independent of dimensionality. We now consider a number of other simplifying features that result from equation~\ref{nu}.

{\bf The monomer overlap concentration}. Returning to the more
usual space fractions, the thermodynamic
properties of a polymer solution above the overlap concentration are determined
by the monomer concentration, $c \sim N_m / L^d$.
In terms of this variable the general form of the dimensionless excess
chemical potential, $\tilde{\mu}^* = \tilde{G}^*/N_p$, valid for all concentrations, is
\begin{equation}
\tilde{\mu}^* = f( c/c* )
\end{equation}
where $c^*$ is the monomer overlap concentration
\begin{equation}
{c^*}^{-1} \sim \frac{L_p^d}{n} \sim n^{d \nu - 1}b^d
\end{equation}
According to equation~\ref{nu} the monomer overlap concentration is ${c^*}^{-1} \sim (n^{1/4} b)^d$.
That is, the inverse monomer overlap concentration is proportional to a length ($n^{1/4}b$),
independent of dimensionality, raised to the power dimensionality.
So, for polymer solutions the excess chemical potential, in terms of monomer concentration,
takes the same form as that of a simple fluid. For the latter, the length
is the range of the potential (for example, the radius of the hard spheres
in a hard sphere fluid).

\begin{figure}[tbp]
\includegraphics[scale=.35]{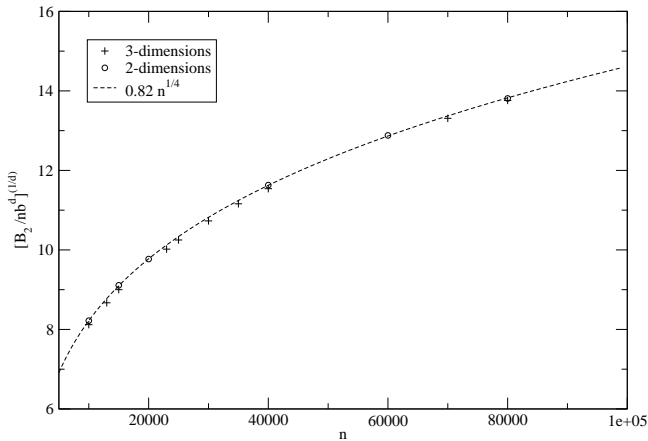}
\caption[a]{Second virial length (defined in the text), in two and three dimensions, as function of the number of steps of the self avoiding walk. The dashed line is a fit to the asymptotic two dimensional result. Data are taken from reference~\cite{Sokal1995}}
\label{B2.fig}
\end{figure}

{\bf The dimensional dependence of the second virial coefficient}.
According to scaling theory the second virial coefficient of
a polymer solution, $B_2$, is proportion to the space occupied by the polymer.
That is $B_2 \propto L_p^d$. From equation~\ref{nu},
we then have $B_2/n \propto b^d n^{d/4}$ so the ``second virial length'' $(B_2/n)^{1/d}$ should
be independent of dimensionality. In figure~\ref{B2.fig} we have plotted this
quantity for the self avoiding random walk in two and three dimensions~\cite{Sokal1995}.
The data for the two dimensionalities are in surprising good quantitative
agreement. There is a small but statistically significant difference for
smaller $n$, but for the largest value ($n=80000$) the difference is
not significant. The reason that this is surprising is because
our argument only predicts that the two functions are proportional.
It appears from the simulation data, at least up to $n=80000$, that
the constant of proportionality is also independent of the dimensionality. If the result for the exponent
given here is {\em approximate}
the two curves shown in the figure must diverge for higher $n$. This could be tested
numerically.

{\bf Hyperbranched polymers}. 
If an ideal chain with $n$ monomers has $m$ branch points, for large $n$ and $m$
its size is reduced by a factor of $m^{1/4}$~\cite{Zimm1949}, relative to the 
linear case, so 
\begin{equation}
\label{zimm}
L_p \sim \frac{n^{1/2}}{m^{1/4}} b
\end{equation}
For a hyperbranched chain ($m \propto n$) we therefore have \mbox{$L_p \sim n^{1/4} b$}. For
all dimensionalities between two and four the hyperbranched ideal chain is an increasingly dense
object. Its size increases more slowly with increasing mass than 
a space filling Euclidean object. This contrasts with the linear chain, which is rarefied
in that its size increases more rapidly with increasing mass. Consequently, it is  generally accepted that the effect
of excluded volume on a hyperbranched chain is to change the sub-Euclidean scaling
to Euclidean scaling, that is \mbox{$L_p \sim n^{1/d}b$}~\cite{Knokolewicz2007}.
On long length scales one expects that correlations can be neglected and that the real hyperbranched chain behaves
like the equivalent hyperbranched ideal chain, but with an $n$ dependent Kuhn length (the ``uniform expansion'' model).
That is, \mbox{$L_p \sim n^{1/2}b_n/m^{1/4}$}, with \mbox{$b_n = n^{\nu(d) - 1/2}b$}. This being the case, we have
\begin{equation}
L_p \sim \frac{n^{\nu(d)}}{m^{1/4}} b
\end{equation}
For the hyperbranched chain we then have \mbox{$L_p \sim n^{\nu(d) - 1/4}$}, so substituting equation~\ref{nu} for the exponent
gives the expected Euclidean scaling, \mbox{$L_p \sim n^{1/d}b$}.

To examine the roles of the microscopic parameters, we
now make use of the ``two parameter model''~\cite{Kosmas1978,Muthukumar1987}. From dimensional
analysis, the size of a single polymer takes the form
\begin{equation}
L_p \sim n^{\frac{1}{2}}b f(z)
\end{equation}
The unknown function $f(z)$
is a function of a dimensionless (but dimensionally
dependent) quantity $z$, 
\begin{equation}
z= \frac{l^d}{b^d} n^{\frac{4-d}{2}}
\end{equation}
Now we have lifted the restriction that we are considering a system with a given value of the ratio $l/b$.

{\bf The microscopic length}. According to scaling theory~\cite{deGennes1979}, for large $z$
the size has a power-law dependence on $n$. That is,
\begin{equation}
\label{two_param}
L_p \sim b^{1-d \gamma(d)} l^{d \gamma(d) }  n^{\frac{1}{2} + \gamma(d)\left( \frac{4-d}{2}  \right)}
\end{equation}
with $\gamma(d)= \left( 2\nu(d) - 1 \right) / \left(4-d \right)$.
So, the dimensional dependence of the scaling exponent also determines the
microscopic length determining the polymer size. For example, the Flory result
gives $l_0 = l^{d/(d+2)} b^{2/(d+2)}$.
Substituting equation~\ref{nu} for the exponent yields \mbox{$\gamma(d)= 1/(2d)$} meaning that
\begin{equation}
L_p \sim l^{\frac{1}{2}} b^{\frac{1}{2}} n^{\nu(d)} 
\end{equation}
There is a pleasing symmetry in the fact that the two 
intrinsic lengths in the problem, $l$ and $b$, enter  with the same power. They play
an equivalent role in determining the size of the polymer. The fact that $l_0$
is independent of dimensionality also means that, as for the ideal chain, the ratio of polymer sizes in two
different dimensionalities is only a function of $n$.

{\bf The excess free energy of dilute polymer solutions}.
Turning to the thermodynamics of low density polymer solutions ($c/c^*<<1$), in the
limit $z \rightarrow 0$ (where the polymer
is hardly expanded) the excess free energy $\tilde{G}^*$ is
\begin{equation}
\tilde{G}^* \sim N_m c l^d
\end{equation}
Thus, the system is thermodynamically equivalent to a simple fluid of monomers with size $l$.
On the other hand, in the scaling ($z \rightarrow \infty$) limit
the excess free energy takes the form
\begin{equation}
\tilde{G}^* \sim N_m c \frac{L_p^d}{n^2}
\end{equation}
In terms of the monomer concentration, using the two parameter expression for $L_p$, yields
\begin{equation}
\label{dilute_fe}
\tilde{G}^* \sim N_m c z^{\left(d \gamma(d) - 1 \right)} l^d
\end{equation}
Introducing the the size of the polymer relative
to the ideal chain size in the dilute limit, $\alpha_0 \sim z^{\gamma(d)}$,
equation~\ref{dilute_fe} becomes
\begin{equation}
\tilde{G}^* \sim N_m c \alpha_0^{\left( \frac{d \gamma(d) - 1}{\gamma(d)} \right)} l^d
\end{equation}
Thus, in the scaling limit the excess free energy of a polymer solution, in terms of the
monomer concentration, takes the same form as that for a simple fluid
except that the interaction length is scaled by a factor related to the degree of
expansion on the polymer. The dependence of this scaling on the dimensionality 
in turn depends on the dimensional dependence of the scaling exponent. For example,
the Flory result gives $\tilde{G}^* \sim N_m c l^d /\alpha_0^2$, independent of dimensionality.
On the other hand, using the expression for $\nu(d)$ given by equation~\ref{nu} gives
\begin{equation}
\label{fe}
\tilde{G}^* \sim N_m c \left( \frac{l}{\alpha_0} \right)^d
\end{equation}
That is,
the system is thermodynamically equivalent to a simple fluid, 
except that the effective size of the
monomers is reduced proportionately to the degree of expansion of the chain, independent of
dimensionality.

\begin{figure}[tbp]
\includegraphics[scale=.35]{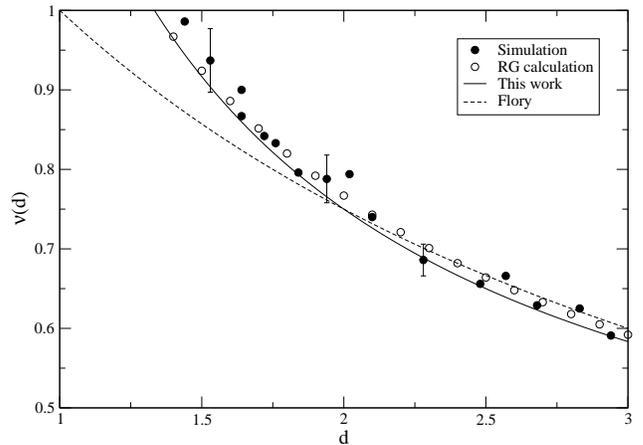}
\caption[a]{Scaling exponent as a function of dimensionality}
\label{cs.fig}
\end{figure}

As noted above, we cannot take one dimension as a point of reference
because in this case the excess free energy, above the overlap concentration, cannot be written in terms 
of both the polymer and monomer concentration.
It is unsurprising, therefore, that equation~\ref{nu} is incorrect for $d=1$. It predicts
$\nu(1)=5/4$. From the condition of fixed contour length,
$\nu(d)$ cannot exceed unity. One interpretation
of this is that the expression predicts a lower critical 
dimensionality at $d=4/3$ (the dimensionality for which the
exponent is unity).
We believe that there is {\em tacit} evidence for this from renormalization
group theory (RG) calculations~\cite{Goettker1999} and simulations in non-integer dimensions~\cite{Alexandrowicz1983}
(see figure~\ref{cs.fig}). We use the word ``tacit'' here because ref.~\cite{Goettker1999} is, to our knowledge,
only published as a pre-print and the author of ref.~\cite{Alexandrowicz1983} interpreted
the low dimensionality behaviour as probably an error in the model because it disagreed
with the Flory result (without explaining
why the same model apparently worked for higher dimensionalities).

Finally we should point out the reasons for believing that the result we
obtain here is not exact.
First, it is outside the range of current numerical and RG estimates for the exponent
in three dimensions. However, the former require extrapolation to the scaling
limit and the latter the re-summation of divergent series. This makes reliably estimating
the error difficult (when Nienhuis argued that $\nu(2)=3/4$,
this was outside the range of contemporaneous numerical and RG values).
Second, equation~\ref{nu} agrees with the epsilon expansion to first
order but {\em not} to second order. Specifically, the epsilon expansion gives
\mbox{$\nu(\epsilon) =1/2[1+\epsilon/8 +15/256 \epsilon^2..]$}, whereas
equation~\ref{nu} gives \mbox{$\nu(\epsilon) =1/2[1+\epsilon/8 +1/32 \epsilon^2..]$}. If
the epsilon expansion is exact and unique to second order then, while equation~\ref{nu} is
correct for $d=3.99$ and $d=2$,
it is only a very good approximation for $d=3$. 
Could the epsilon expansion to second order be inexact or non-unique? 
It seems heresy to even suggest this, but we note that
there is an added degree of subtlety in calculating the second order
term in the epsilon expansion as compared to the first order term. Namely, a ``magic interaction
strength'' is required for which corrections to scaling disappear~\cite{Wilson1972b}. 
Further, the epsilon expansion by its nature
requires the concept of non-integer spacial dimensions. Are non-integer dimensions
unique or just an interpolation between integer dimensions? In the analysis above
we have treated $d$ as a continuous variable, but this is not actually necessary. We could equally
well have restricted ourselves to integer dimensionalities.  The concept of non-integer dimensionality
is not required here.
To summarize, there are two possibilities. The epsilon expansion is exact and unique.
All the above simplifications apply in $4,3.99$ and $2$ dimensions, but not {\em quite} in
three. 
Alternatively, nature is kind and there is nothing
special about three dimensions.

We would like to thank Daan Frenkel and Bela Mulder for their interest and encouragement,
and participants of the Frenkel weekend for their vote of confidence.

\end{document}